\documentclass[useAMS,usenatbib]{mn2e}

\usepackage{graphics,epsfig,psfig}
\usepackage{graphicx}
\usepackage{amssymb}
\title[XMD galaxies]{When are extremely metal-deficient galaxies extremely metal-deficient?}
\author[]
{B. Ekta,$^1$\thanks{ekta@ncra.tifr.res.in}
Jayaram N. Chengalur$^1$\\
$^1$ National Centre for Radio Astrophysics, Post Bag 3, Ganeshkhind, Pune 411 007, India}

\DeclareRobustCommand{\ion}[2]{%
\relax\ifmmode
\ifx\testbx\f@series
{\mathbf{#1\,\mathsc{#2}}}\else
{\mathrm{#1\,\mathsc{#2}}}\fi
\else\textup{#1\,{\mdseries\textsc{#2}}}%
\fi}

\newcommand{\kms}{km~s$^{-1}$}
\newcommand{\HI}{\ion{H}{i}}
\newcommand{\HII}{\ion{H}{ii}}

\newcommand{\peff}{\ensuremath{p_{\rm eff}}}
\newcommand{\fgas}{\ensuremath{\mu_{\rm gas}}}
\begin{document}
 
\label{firstpage}

\date{Accepted 2010 Received 2009}

\pagerange{\pageref{firstpage}--\pageref{lastpage}} \pubyear{2010} 

\maketitle

\begin{abstract}

Extremely metal-deficient (XMD) galaxies, by definition, have oxygen abundances 
$\le$~1/10 solar, and form a very small fraction of the local gas-rich, 
star-forming dwarf galaxy population. We examine their positions in the 
luminousity--metallicity (L--Z) and mass-metallicity (M--Z) planes, with 
respect to the L--Z and M--Z relations of other gas-rich, star-forming
dwarf galaxies, viz., blue compact galaxies (BCGs) and dwarf irregular (dI) 
galaxies.
We find that while the metallicities of some low-luminousity XMD galaxies are
consistent with those expected from the L--Z relation, other XMD galaxies 
are deviant, and more so as the luminousity and/or metal-deficiency increases.
We determine the 95 per cent confidence interval around the L--Z relation 
for BCGs, and find that its lower boundary is given by  
12~+~log(O/H)~=~--0.177~${\rm M}_{\rm B}$~+~4.87. We suggest that a galaxy 
should be regarded as XMD, in a statistically significant manner, only if 
it lies below this boundary in the L--Z plane. Of our sample of XMD 
galaxies, we find that more than half are XMD by this criterion, and in 
fact, nine of the galaxies lie below the 99.5 per cent confidence interval 
about the L--Z relation.

We also determine the gas mass fractions and chemical yields of galaxies in 
all three samples. We find that the effective chemical yield increases 
with increasing baryonic mass, consistent with what is expected if outflows of 
metal-enriched gas are important in determining the effective yield. XMD 
galaxies have lower effective yield than BCG/dI galaxies of similar baryonic
mass. This suggests that some process, peculiar to XMD galaxies, has resulted 
in their low measured metallicities. Motivated by the fact that interactions 
are common in XMD galaxies, we suggest that improved (tidally-driven) mixing of 
the interstellar media (ISM) in XMD galaxies leads to a lowering of both, the 
measured metallicity and the calculated effective yield. In isolated dwarf 
galaxies, the outer parts of the stellar envelope probably do not participate 
in the star formation, but are still generally included in the calculation of 
effective yield. This results in an overestimate of the effective yield. 
We suggest that XMD galaxies are deviant from the L--Z relation because of a 
combination of being gas-rich (i.e., having processed 
less gas into stars) and having more uniform mixing of metals in their ISM.
\end{abstract}


\section{INTRODUCTION}
\label{sec:intro}

Extremely metal-deficient (XMD) galaxies are defined as galaxies with 
interstellar medium (ISM) oxygen abundances less than or equal to 
0.1~Z$_{\odot}$, i.e., 12~+~log(O/H)~$\le$~7.65 (e.g., Kunth \& Ostlin 2000; 
Kniazev et al. 2003; Pustilnik \& Martin 2007; Brown, Kewley \& Geller 
2008). Such metallicities are more typical of primeval galaxies at high 
red-shifts, and it is not firmly established 
that why such low-metallicity star-forming galaxies exist in the 
local Universe. Their low metallicities could be due to (1)~slow chemical 
evolution, i.e., a lower than usual time-averaged star formation rate, 
(2)~infall of 
metal-poor gas, (3)~preferential loss of metal-enriched gas through galactic 
winds. While XMD galaxies form a very small fraction of the gas-rich, 
star-forming dwarf galaxy population (i.e., blue compact and 
dwarf irregular galaxies), the mere existence of a local population 
of chemically-unevolved galaxies is noteworthy. Do these galaxies 
represent a separate population, or are they, merely, the low-metallicity 
tail of the metallicity distribution of galaxies? 

In trying to address this question, a useful starting point is to note that
for galaxies, in general, metallicity correlates with both luminousity and mass 
(e.g., Skillman, Kennicutt \& Hodge 1989; Tremonti et al. 2004; Lee et al. 
2006). Possible reasons for such a correlation are (1)~that the star formation 
efficiency increases with mass, so that low-mass galaxies end up converting a 
smaller fraction of their mass into stars, and hence having lower 
metallicities, or 
(2)~that massive galaxies, by virtue of having deeper potential wells,
are more capable of retaining their metals instead of losing them through 
enriched galactic winds. The relative importance of these two factors 
is not well established. For example, Brooks et al. (2007) show that star 
formation efficiencies, regulated by stellar feedback, lead to low  
metallicity for low-mass galaxies. On the other hand, Tremonti et al. (2004), 
from a study of star-forming Sloan Digital Sky Survey (SDSS) galaxies, find 
evidence in favour of the correlation being driven by decreasing metal-loss 
efficiency with increasing mass [but, see also Lee et al. (2006) and Vale Asari 
et al. (2009), who argue against this]. In any case, since galaxy metallicity 
is known to correlate with both the luminousity and mass, the question 
regarding the nature of XMD galaxies is that are these galaxies outliers with 
respect to the luminousity--metallicity (L--Z) or mass--metallicity (M--Z)
correlations? Similarly, does a comparison of the effective chemical yields of 
XMD and other dwarf galaxies give some clue as to why XMD galaxies are 
metal-poor?

  To answer these questions, we have constructed a sample of XMD galaxies,
along with comparison samples of the two major classes of gas-rich, 
star-forming dwarf galaxies, viz., blue compact galaxies (BCGs) and dwarf 
irregular (dI) galaxies. We derive L--Z and M--Z correlations 
for each of the BCG and dI samples, and see where XMD galaxies lie with 
respect to these correlations. We also calculate and compare their 
effective chemical yields with BCGs and dI galaxies. To the best of our 
knowledge, this is the first derivation of the M--Z relations for BCG 
galaxies, as a class.

Our paper is divided into the following sections. We describe our samples in 
Section~\ref{sec:sam}. The derived L--Z and M--Z relations and chemical 
yields are given in Section~\ref{sec:lzmz}. We discuss and summarise our 
results in Section~\ref{sec:dis}. 

\section{SAMPLES} 
\label{sec:sam}

We have selected a sample of 59 local ($V_{\rm hel} <$ 4500 km~s$^{-1}$) BCGs 
from emission-line galaxies in the SDSS Data release~3 (DR3), whose abundance 
measurements are published by Izotov et al. (2006a). We have chosen 
galaxies with compact optical morphology, 
as seen in their {\it g}-band SDSS images, for our BCG sample. We have no bias 
towards metallicity in our selection, except that we have chosen only those 
galaxies for which oxygen abundances, given by 12~+~log(O/H), have been 
determined to within 0.1~dex. However, very few BCGs were dropped due to 
this cut-off on accuracy of determined metallicities, and we confirmed that 
it is not introducing any luminousity bias in our sample. Also, note that the 
majority of the galaxies have much 
more accurate determinations of oxygen abundances than this. Four of these BCGs 
have metallicities that fall in the XMD regime. 
We have calculated the {\it B} and {\it R-}band magnitudes of the galaxies 
from their SDSS {\it g} and {\it r} magnitudes, using 
the formulae derived by Lupton (2005), which are given on the SDSS web-site. 
The parameters of galaxies, in our BCG sample, are given in 
Table~\ref{tab:parbcg}. 

We have searched the literature for dI galaxies with known oxygen 
abundances and optical magnitudes, and selected 32 galaxies for representative 
dI sample. All of them are well-studied, and have 
been classified as dIs by other researchers and/or NASA/IPAC Extragalactic 
Database (NED). 
Relevant parameters for this sample are given in Table~\ref{tab:pardi}. Eight 
of these dIs have metallicities that fall in the XMD regime. For XMD 
galaxies also, we searched the literature, and chose known nearby 
(v$_{hel}$~$<$4500~\kms) XMD galaxies for our sample. Table~\ref{tab:parxmd} 
gives the corresponding parameters for the 31 XMD galaxies in our sample. 
Oxygen abundances of all BCGs, XMD and dI galaxies, except UGC~12613, UGC~4117, 
are determined by the more accurate `direct' method.     

\begin{table*}
\begin{tabular}{lccccc}
\hline
Name & m$_{B}$ & 12~+~log(O/H) & log(N/O) & \HI\ flux & References \\
    &         &   &     & (Jy~kms$^{-1}$) & \\
\hline
SDSS~J002425.94+140410.3 & 16.41 & 8.39~$\pm$~0.07 & --1.48~$\pm$~0.09 &  & \\
SDSS~J005319.63--102411.8 & 16.60 & 8.14~$\pm$~0.10 & --1.00~$\pm$~0.13 &  & \\
SDSS~J011914.27--093546.4 & 17.88 & 7.61~$\pm$~0.04 & --1.38~$\pm$~0.07 & 1.5 & 1 \\
UM~323 & 16.16 & 7.96~$\pm$~0.04 & --1.50~$\pm$~0.06 & 2.4 & 2 \\
SDSS~J021852.90--091218.7 & 18.67 & 7.89~$\pm$~0.03 & --1.49~$\pm$~0.06 & & \\
SDSS~J023145.99--090847.6 & 17.05 & 8.16~$\pm$~0.03 & --1.54~$\pm$~0.05 & & \\
SDSS~J024815.93--081716.5 & 16.04 & 7.98~$\pm$~0.02 & --1.55~$\pm$~0.03 & & \\
SDSS~J025346.36--072343.6 & 16.75 & 7.89~$\pm$~0.03 & --1.48~$\pm$~0.04 & 1.5 & 1 \\
SDSS~J025426.12--004122.6 & 17.61 & 8.06~$\pm$~0.05 & --1.46~$\pm$~0.07 & & \\
SDSS~J082555.52+353231.9 & 17.93 & 7.42~$\pm$~0.03 & --1.50~$\pm$~0.07 & 0.3 & 3 \\
SDSS~J082718.07+460203.0 & 15.79 & 8.23~$\pm$~0.07 & --1.42~$\pm$~0.09 & &  \\
MRK~0627 & 16.01 & 8.24~$\pm$~0.04 & --1.34~$\pm$~0.06 & & \\
SDSS~J085920.83+005142.1 & 17.26 & 8.09~$\pm$~0.06 & --1.32~$\pm$~0.08 & & \\
SDSS~J091028.77+071117.9 & 16.67 & 7.63~$\pm$~0.03 & --1.41~$\pm$~0.05 & 2.0 & 4 \\
UGCA~154 & 15.18 & 8.22~$\pm$~0.07 & --1.23~$\pm$~0.10 & & \\
MCG~+09--15--115 & 16.08 & 8.12~$\pm$~0.05 & --1.28~$\pm$~0.06 & & \\
MRK~1416 & 16.32 & 7.89~$\pm$~0.02 & --1.50~$\pm$~0.03 & 2.0 & 5 \\  
SBS~0926+606A & 16.22 & 7.99~$\pm$~0.02 & --1.46~$\pm$~0.04 & 1.3 & 6 \\
UGCA~184 & 16.11 & 8.04~$\pm$~0.02 & --1.44~$\pm$~0.03 & 2.2 & 5 \\
SDSS~J095241.42+020758.1 & 17.71 & 7.92~$\pm$~0.04 & --1.42~$\pm$~0.06 & & \\
UGCA~208 & 15.35 & 8.30~$\pm$~0.08 & --1.28~$\pm$~0.11 & & \\
SDSS~J103137.27+043422.0 & 16.40 & 7.70~$\pm$~0.04 & --1.35~$\pm$~0.11 & & \\
MRK~1434 & 16.59 & 7.79~$\pm$~0.03 & --1.51~$\pm$~0.06 & & \\
SBS~1037+494 & 16.93 & 8.01~$\pm$~0.03 & --1.49~$\pm$~0.04 & 1.9 & 7 \\
SDSS~J104456.00+004433.4 & 16.79 & 7.82~$\pm$~0.03 & --1.45~$\pm$~0.05 & & \\
CGCG~010--041 & 15.83 & 8.38~$\pm$~0.10 & --1.39~$\pm$~0.13 & 5.4 & 8 \\
SDSS~J105741.94+653539.7 & 16.70 & 8.29~$\pm$~0.04 & --1.45~$\pm$~0.06 & & \\
TOLOLO~1108+098 & 15.47 & 8.15~$\pm$~0.08 & --1.46~$\pm$~0.11 & 2.8 & 2 \\
SDSS~J112526.75+654607.2 & 17.88 & 7.81~$\pm$~0.05 & --1.60~$\pm$~0.08 & & \\
MRK~1446 & 16.07 & 8.13~$\pm$~0.04 & --1.51~$\pm$~0.06 & 0.9 & 7 \\
SDSS~J112742.97+641001.4 & 17.52 & 8.18~$\pm$~0.10 & --1.36~$\pm$~0.13 & & \\
SDSS~J113341.19+634925.9 & 17.59 & 7.97~$\pm$~0.04 & --1.49~$\pm$~0.06 & & \\
UM~454 & 16.40 & 7.94~$\pm$~0.07 & --1.28~$\pm$~0.09 & 0.8 & 2 \\
UM~461 & 16.05 & 7.81~$\pm$~0.02 & & 3.0 & 2 \\
UM~462 & 14.45 & 7.82~$\pm$~0.02 & --1.49~$\pm$~0.03 & 5.6 & 2 \\
UM~463 & 17.66 & 7.81~$\pm$~0.03 & --1.49~$\pm$~0.05 &  & \\
SBS~1205+557 & 17.26 & 7.92~$\pm$~0.04 & --1.53~$\pm$~0.06 & 0.2 & 7 \\
MRK~1313 & 16.11 & 8.19~$\pm$~0.06 & --1.46~$\pm$~0.08 & 1.6 & 2 \\
SBS~1211+540 & 17.30 & 7.65~$\pm$~0.03 & --1.58~$\pm$~0.08 & 0.6 & 7 \\
HARO~06 & 15.21 & 8.25~$\pm$~0.07 & --1.29~$\pm$~0.08 & 2.1 & 9 \\
SBS~1222+614 & 14.95 & 8.00~$\pm$~0.02 & --1.55~$\pm$~0.04 & 2.8 & 10 \\
VCC~1744 & 17.12 & 7.82~$\pm$~0.03 & --1.51~$\pm$~0.05 & & \\
SDSS~J124159.34--034002.4 & 18.47 & 7.68~$\pm$~0.04 & --1.39~$\pm$~0.06 & & \\
UM~538 & 17.51 & 8.02~$\pm$~0.04 & --1.42~$\pm$~0.06 & 0.2 & 9 \\
MRK~1480 & 16.31 & 8.04~$\pm$~0.05 & --1.44~$\pm$~0.07 & 2.8 & 7 \\
SDSS~J135030.82+622649.0 & 17.44 & 7.91~$\pm$~0.05 & --1.41~$\pm$~0.07 & & \\
UM~618 & 17.92 & 7.97~$\pm$~0.06 & --1.48~$\pm$~0.08 & 0.2 & 9 \\
SBS~1423+517 & 16.61 & 8.06~$\pm$~0.05 & --1.34~$\pm$~0.06 & 0.9 & 7 \\
SBS~1428+457 & 15.50 & 8.40~$\pm$~0.05 & --1.49~$\pm$~0.06 & 4.1 & 7 \\
SDSS~J143053.51+002746.2 & 16.67 & 8.15~$\pm$~0.03 & --1.61~$\pm$~0.05 &  & \\
SBS~1453+526 & 17.13 & 8.02~$\pm$~0.07 & --1.35~$\pm$~0.09 & 0.7 & 7 \\
SDSS~J153704.18+551550 & 15.50 & 8.08~$\pm$~0.02 & --1.61~$\pm$~0.03 &  & \\
SDSS~J161623.53+470202.3 & 16.69 & 7.98~$\pm$~0.04 & --1.56~$\pm$~0.06 &  & \\
SDSS~J165730.29+384122.9 & 17.03 & 7.94~$\pm$~0.04 & --1.54~$\pm$~0.06 &  & \\
SDSS~J171236.63+321633.4 & 17.95 & 7.89~$\pm$~0.03 & --1.55~$\pm$~0.06 &  & \\
MRK~0894 & 15.67 & 8.44~$\pm$~0.10 & --1.14~$\pm$~0.13 &  & \\
SDSS~J211942.38--073224.3 & 17.60 & 7.98~$\pm$~0.05 & --1.43~$\pm$~0.07 & 8.4 & 8 \\
SDSS~J223036.79--000636.9 & 17.38 & 7.66~$\pm$~0.04 &  & 1.3 & 1 \\
UM~158 & 16.90 & 8.10~$\pm$~0.05 & --1.40~$\pm$~0.07 &  & \\
\hline
\end{tabular}
\caption{Parameters of galaxies in the BCG sample. {\bf References} (for \HI\ fluxes): 
1. Geha et al. (2006); 2. Smoker et al. (2000); 3. Chengalur et al. (2006); 
4. Hogg et al. (1998); 5. Thuan et al. (1999); 6. Pustilnik et al. (2002); 
7. Huchtmeier, Krishna \& Petrosian (2005); 8. Doyle et al. (2005); 9. Salzer et al. 
(2002); 10. Huchtmeier et al. (2007)}
\label{tab:parbcg}
\end{table*}

\begin{table*}
\begin{tabular}{lccccc}
\hline
Name & m$_{B}$ & 12~+~log(O/H) & log(N/O) & \HI\ flux & References \\
    &         &   &     & (Jy~kms$^{-1}$) &  \\ 
\hline
UGC~12894 & 16.55 & 7.56~$\pm$~0.04 & --1.54~$\pm$~0.08 & 4.5 & 1, 2, 3 \\
WLM & 11.30 & 7.83~$\pm$~0.06 & --1.49~$\pm$~0.08 & 294.0 & 4, 5, 6 \\ 
UGC~00300 & 15.86 & 7.80~$\pm$~0.03 & --1.50~$\pm$~0.11 & 4.5 & 7, 7, 8 \\
ESO~473--~G~024 & 16.04 & 7.45~$\pm$~0.03 & --1.43~$\pm$~0.03 & 5.7 & 9, 10, 11 \\
IC~1613 & 10.25 & 7.62~$\pm$~0.05 & --1.13~$\pm$~0.18 & 482.0 & 4, 12, 13 \\
UGC~00685 & 14.20 & 8.00~$\pm$~0.03 & --1.45~$\pm$~0.08 & 11.8 & 1, 12, 14 \\
UGC~01104 & 14.41 & 7.94~$\pm$~0.04 & --1.67~$\pm$~0.07 & 11.2 & 1, 2, 15 \\
UGCA~020 & 15.78 & 7.58~$\pm$~0.03 & --1.57~$\pm$~0.11 & 10.2 & 16, 16, 16\\
ESO~245--~G~005 & 12.60 & 7.70~$\pm$~0.10 & --1.27~$\pm$~0.10 & 87.3 & 17, 12, 18 \\
UGC~02023 & 13.88 & 8.01~$\pm$~0.02 & --1.35~$\pm$~0.06 & 18.7 & 1, 2, 19 \\
UGC~03174 & 15.02 & 7.83~$\pm$~0.09 &  & 18.6 & 7, 7, 8 \\
ESO~489--~G?056 & 15.70 & 7.49~$\pm$~0.10 & --1.35~$\pm$~0.20 & 2.8 & 9, 12, 20 \\
UGC~03647 & 14.67 & 8.07~$\pm$~0.04 & --1.28~$\pm$~0.07 &  & 1, 2 \\
UGC~03672 & 15.40 & 8.01~$\pm$~0.04 & --1.64~$\pm$~0.12 & 15.4 & 1, 2, 8 \\
UGC~03974 & 13.62 & 7.92~$\pm$~0.06 &  & 61.0 & 21, 22, 23 \\
UGC~4117 & 15.34 & 7.89~$\pm$~0.10 & --1.52~$\pm$~0.15 & 4.1 & 1, 2, 24 \\
Holmberg~II & 11.54 & 7.71~$\pm$~0.13 &  & 267.0 & 4, 25, 26 \\
UGC~04483 & 15.29 & 7.56~$\pm$~0.03 & --1.57~$\pm$~0.07 & 13.6 & 27, 12, 28 \\
Sextans~B & 12.07 & 7.84~$\pm$~0.05 & --1.46~$\pm$~0.06 & 102.4 & 4, 29, 14 \\
UGC~05423 & 14.42 & 7.98~$\pm$~0.10 & --1.26~$\pm$~0.25 & 3.8 & 21, 30, 31 \\
NGC~4214 & 10.32 & 8.22~$\pm$~0.05 & --1.30~$\pm$~0.15 & 261.7 & 4, 32, 19 \\
UGCA~292 & 15.86 & 7.30~$\pm$~0.03 & --1.45~$\pm$~0.07 & 17.6 & 4, 12, 33 \\
NGC~4789A & 14.20 & 7.67~$\pm$~0.06 & --1.68~$\pm$~0.13 & 105.0 & 4, 12, 34 \\ 
UGC~08091 & 14.70 & 7.65~$\pm$~0.06 & --1.51~$\pm$~0.07 & 9.7 & 4, 12, 33 \\ 
UGCA~357 & 15.63 & 8.04~$\pm$~0.04 & --1.55~$\pm$~0.14 & 13.2 & 7, 7, 8 \\
UGC~08651 & 14.45 & 7.85~$\pm$~0.04 & --1.60~$\pm$~0.09 & 11.4 & 35, 12, 28 \\ 
UGC~09128 & 14.46 & 7.75~$\pm$~0.05 & --1.80~$\pm$~0.12 & 13.9 & 1, 12, 19 \\ 
UGC~09240 & 13.31 & 7.95~$\pm$~0.03 & --1.60~$\pm$~0.06 & 24.6 & 1, 12, 36 \\ 
UGC~9992 & 14.86 & 7.88~$\pm$~0.12 & --1.26~$\pm$~0.19 & 11.8 & 1, 2, 19 \\ 
NGC~6822 & 9.81 & 8.11~$\pm$~0.05 & --1.60~$\pm$~0.10 & 2266.0 & 4, 12, 37 \\ 
UGC~12613 & 12.50 & 7.93~$\pm$~0.14 & --1.24~$\pm$~0.15 &  & 1, 12 \\
UGC~12713 & 14.91 & 7.80~$\pm$~0.06 & --1.54~$\pm$~0.11 & 10.8 & 1, 2, 38 \\
\hline
\end{tabular}
\caption{Parameters of galaxies in the dI sample. {\bf References}  
(m$_{B}$, 12~+~log(O/H) and log(N/O), \HI\ fluxes): 1. van Zee (2000); 
2. van Zee \& Haynes (2006); 3. Staveley-Smith, Davies \& Kinman 
(1992); 4. Hunter \& Elmegreen (2006); 5. Lee, Skillman \& Venn 
(2005); 6. Kepley et al. (2007); 7. van Zee, Haynes \& Salzer 
(1997); 8. van Zee et al. (1997a), 9. Parodi, Barazza \& Binggeli 
(2002); 10. Skillman, Cote \& Miller (2003); 11. Warren, Jerjen \& 
Koribalski (2006); 12. van Zee, Skillman \& Haynes (2006); 13. 
Silich et al. (2006); 14. Hoffman et al. (1996); 15. Smoker 
et al. (2000); 16. van Zee et al. (1996); 17. Makarova 
et al. (2005); 18. Cote et al. (1997); 19. Swaters et al. 
(2002); 20. Pustilnik \& Martin (2007); 21. Makarova (1999); 
22. Lee, Zucker \& Grebel (2007); 23. Walter \& Brinks (2001); 
24 O'Neil (2004); 25. Lee et al. (2003); 26. Bureau \& Carignan 
(2002); 27. Gil de Paz, Madore \& Pevunova (2003); 28. Huchtmeier, 
Karachentsev \& Karachentseva (2003); 29. Kniazev et al. (2005); 
30. Miller \& Hodge (1996); 31. Walter et al. (2007); 32. Lee 
et al. (2006); 33. Young et al. (2003); 34. Carignan \& Purton (1998); 
35. Makarova et al. (1998); 36. Springob et al. (2005); 37. 
de Blok \& Walter (2006); 38. Noordermeer et al. (2005).}
\label{tab:pardi}
\end{table*}

\begin{table*}
\begin{tabular}{lccccc}
\hline
Name & m$_{B}$ & 12~+~log(O/H) & log(N/O) & \HI\ flux & References \\
    &         &   &     & (Jy~kms$^{-1}$) & \\
\hline
UGC~772 & 13.73 & 7.24~$\pm$~0.05 &  & 5.3 & 1,2,3 \\
HS~0122+0743 & 15.50 & 7.60 &  & 7.6 & 4,5,4 \\  
SDSS~J0133+1342 & 18.42 & 7.60~$\pm$~0.03 &  & 0.10 & 6,7,4 \\
UM~133 & 15.57 & 7.63~$\pm$~0.02 &  & 4.2 & 8,9,10 \\ 
KUG~0201--103 & 17.35 & 7.56~$\pm$~0.03 &  &  & 6,2 \\ 
KUG~0203--100 & 15.15 & 7.61~$\pm$~0.09 &  & 12.2 & 6,7,4 \\
SDSS~J030149.02--005257.3 & 18.64 & 7.52~$\pm$~0.06 &  & & 6,2 \\ 
SBS~0335--052W & 19.14 & 7.12~$\pm$~0.03 & --1.60~$\pm$~0.04 & 0.86 & 11,12,12,13 \\ 
SBS~0335--052E & 16.95 & 7.29~$\pm$~0.02 & --1.53~$\pm$~0.03 & 0.61 & 11,14,14,13 \\
HS~0846+3522 & 18.19 & 7.65 &  & 0.10 & 6,4,4 \\ 
SDSS~J091159.43+313535.9 & 18.01 & 7.51~$\pm$~0.14 &  &  & 6,2 \\ 
I~Zw~18 & 16.06 & 7.25~$\pm$~0.04 & --1.64~$\pm$~0.07 & 2.97 & 15,16,16,17 \\ 
KUG~0937+298 & 16.19 & 7.65~$\pm$~0.04 & --1.49 & 2.05 & 6,18,18,4 \\ 
SBS~0940+544 & 17.23 & 7.43~$\pm$~0.01 & --1.61~$\pm$~0.03 &  & 8,19,19 \\
DDO~68 & 14.60 & 7.14~$\pm$~0.03 &  & 28.9 & 20,2,3 \\ 
HS~1013+3809 & 15.99 & 7.59 &  & 1.51 & 6,4,4 \\ 
HS~1033+4757 & 17.76 & 7.65 &  & 1.32 & 6,4,4 \\ 
SDSS~J1044+0353 & 17.39 & 7.46~$\pm$~0.03 & --1.69~$\pm$~0.08 &  & 6,16,16 \\
HS~1059+3934 & 17.63 & 7.62 &  & 1.39 & 6,4,4 \\
SDSS~J1105+6022 & 16.46 & 7.64~$\pm$~0.04 &  & 2.48 & 6,7,4 \\
SBS~1116+517 & 17.13 & 7.51~$\pm$~0.04 &  & 1.31 & 6,7,21 \\
SDSS~J1121+0324 & 16.89 & 7.64~$\pm$~0.08 &  & 2.67 & 6,7,4 \\
SBS~1129+576 & 16.40 & 7.41~$\pm$~0.08 & --1.52 & 3.9 & 6,22,22,23 \\
SBS~1159+545 & 18.39 & 7.49~$\pm$~0.01 & --1.58~$\pm$~0.04 &  & 6,19,19 \\
Tol~1223--359 & 17.58 & 7.54~$\pm$~0.01 & --1.60~$\pm$~0.02 & 2.13 & 8,24,24,4 \\
KISSR~1490 & 19.34 & 7.56~$\pm$~0.07 & --1.41 &  & 6,25,25 \\
SBS~1415+437 & 15.47 & 7.61~$\pm$~0.01 & --1.56~$\pm$~0.02 & 5.40 & 8,26,26,21 \\
HS~1442+4250 & 15.61 & 7.63~$\pm$~0.02 & --1.44~$\pm$~0.04 & 7.05 & 8,27,27,4\\
HS~1704+4332 & 18.41 & 7.55 &  & 0.24 & 8,4,4 \\
SDSS~J210455.31-003522.2 & 18.04 & 7.26~$\pm$~0.03 &  & 2.0 & 6,28,3 \\
HS~2134+0400 & 19.30 & 7.44 &  & 0.12 & 29,29,4 \\
\hline
\end{tabular}
\caption{Parameters of the studied XMD galaxies. {\bf References}: 
1. Patterson \& Thuan (1996); 2. Izotov \& Thuan (2007); 3. Ekta, 
Chengalur \& Pustilnik (2008); 
4. Pustilnik \& Martin (2007); 5. Ugryumov et al. (2003); 6. SDSS; 
7. Kniazev et al. (2003); 8. Gil de Paz et al. (2003); 9. Kniazev et al. (2001); 
10. Smoker et al. (2000); 11. Pustilnik, Pramskij \& Kniazev (2004); 12. Izotov, Thuan \& Guseva (2005); 
13. Ekta, Pustilnik \& Chengalur (2009); 14. Izotov et al. (2006c); 
15. Papaderos et al. (2002); 16. Izotov et al. (2006a); 17. van Zee et al. (1998); 
18. Lee, Salzer \& Melbourne (2004); 19. Izotov \& Thuan (1999); 
20. Pustilnik, Kniazev \& Pramskij (2005); 21. Huchtmeier et al. (2005); 
22. Guseva et al. (2003a); 23. Ekta, Chengalur \& Pustilnik (2006); 
24. Izotov et al. (2004); 25. Melbourne et al. (2004); 26. Guseva et al. (2003c); 
27. Guseva et al. (2003b); 28. Izotov et al. (2006b); 29. Pustilnik et al. (2006)}
\label{tab:parxmd}
\end{table*}

Stellar masses were computed using formulae given by Bell \& de Jong (2001), 
using their formation epoch: bursts model. The mass-to-luminousity ratio 
was computed using extinction-corrected {\it B-V} 
colours, whenever available. For galaxies where the {\it B-V} colour was not 
available, we used extinction-corrected {\it B-R} colours. For galaxies 
for which both {\it B-V} and {\it B-R} colours are available, ratios found 
from {\it B-V} colour were systematically higher than those obtained from 
{\it B-R} colour by a factor of $\sim$1.38. While this is within the
variations in the model mass-to-luminousity ratio given by  Bell \& de Jong 
(2001), we
have none the less, for consistency, scaled the stellar masses obtained 
from the {\it B-R} by 1.38. Since CO emission is not detected, or detected at a very low level, in most 
of the dwarf galaxies (e.g., Taylor, Kobulnicky \& Skillman 1998), we have 
neglected the contribution of molecular gas for our samples, assuming that 
it is small. We have, however, scaled \HI\ masses by a factor of 1.33, to 
account for He.

Following Lee et al. (2003) and Vaduvescu, McCall \& Richer (2007), the linear fits 
for the L--Z and M--Z correlations, of each of the dI and BCG samples, were 
obtained using geometric mean functional relationship, so that a fair 
comparison between relations derived here and those derived in these earlier 
works can be drawn. This method assumes similar dispersions in both 
observables, and is apt for use when there are uncertainties in measurements 
of both variables.

\section{RESULTS}
\label{sec:lzmz}

\begin{figure*}
\includegraphics[width=6.15cm,angle=270]{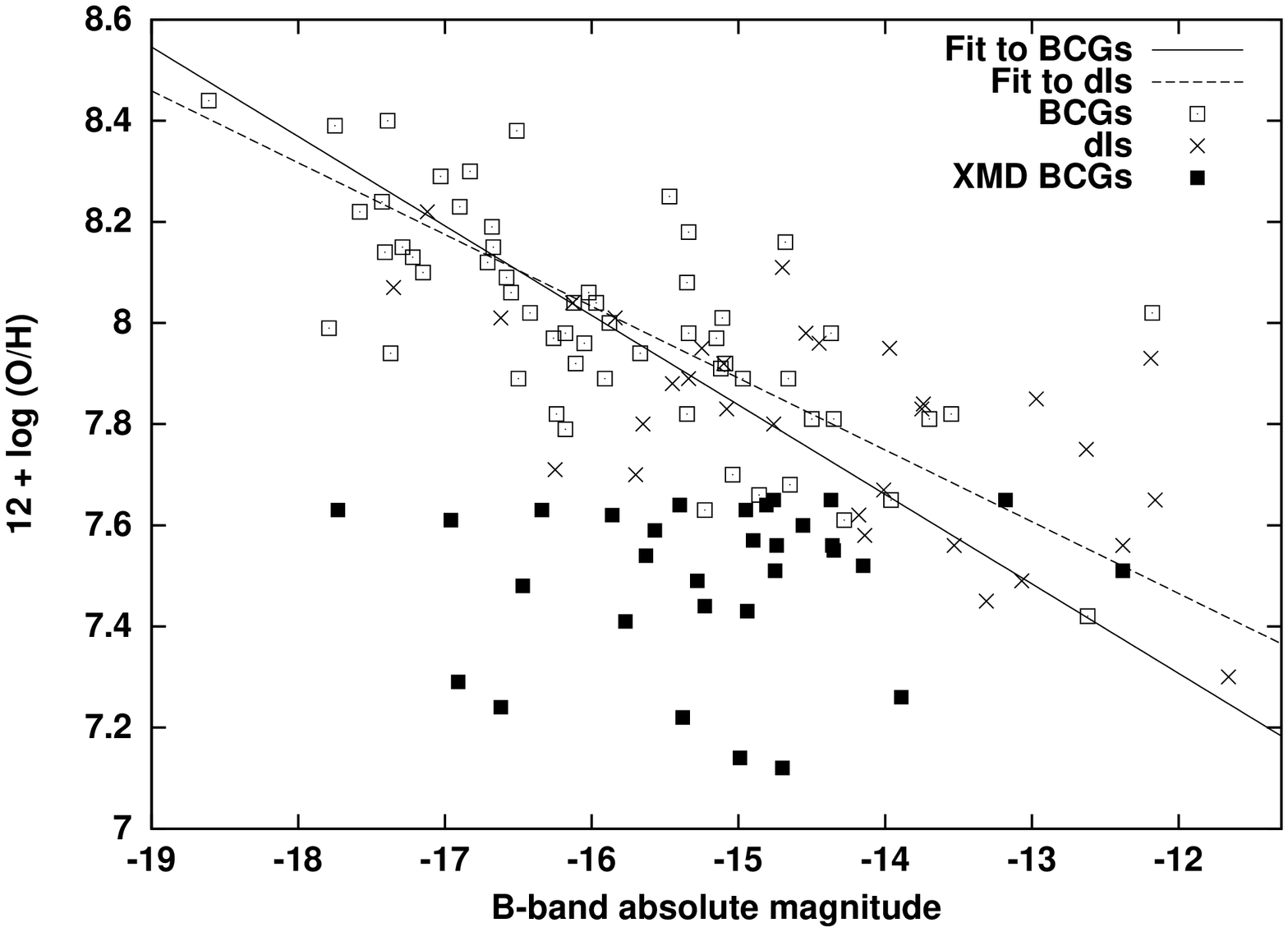}
\includegraphics[width=6.15cm,angle=270]{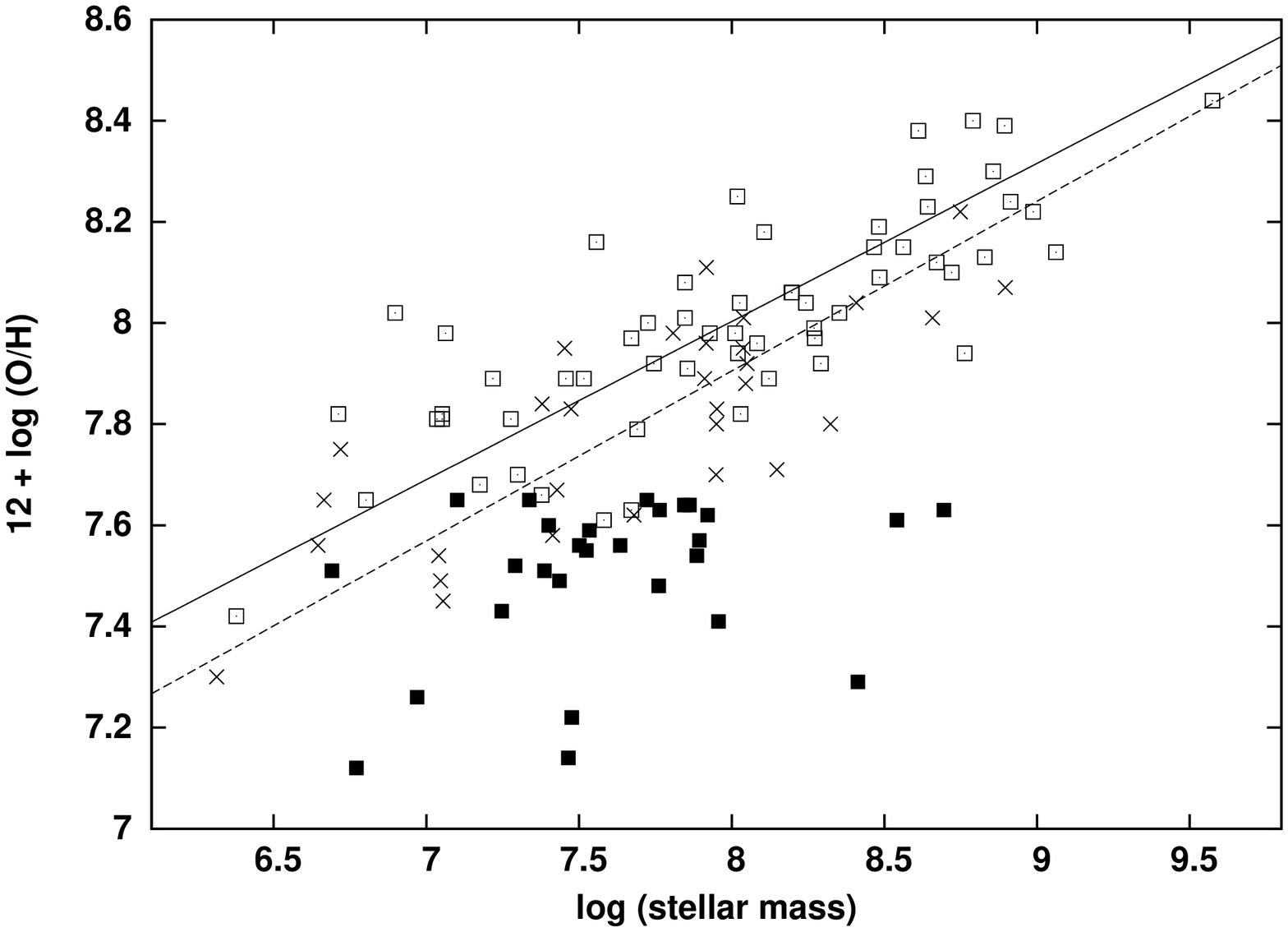}
\includegraphics[width=6.15cm,angle=270]{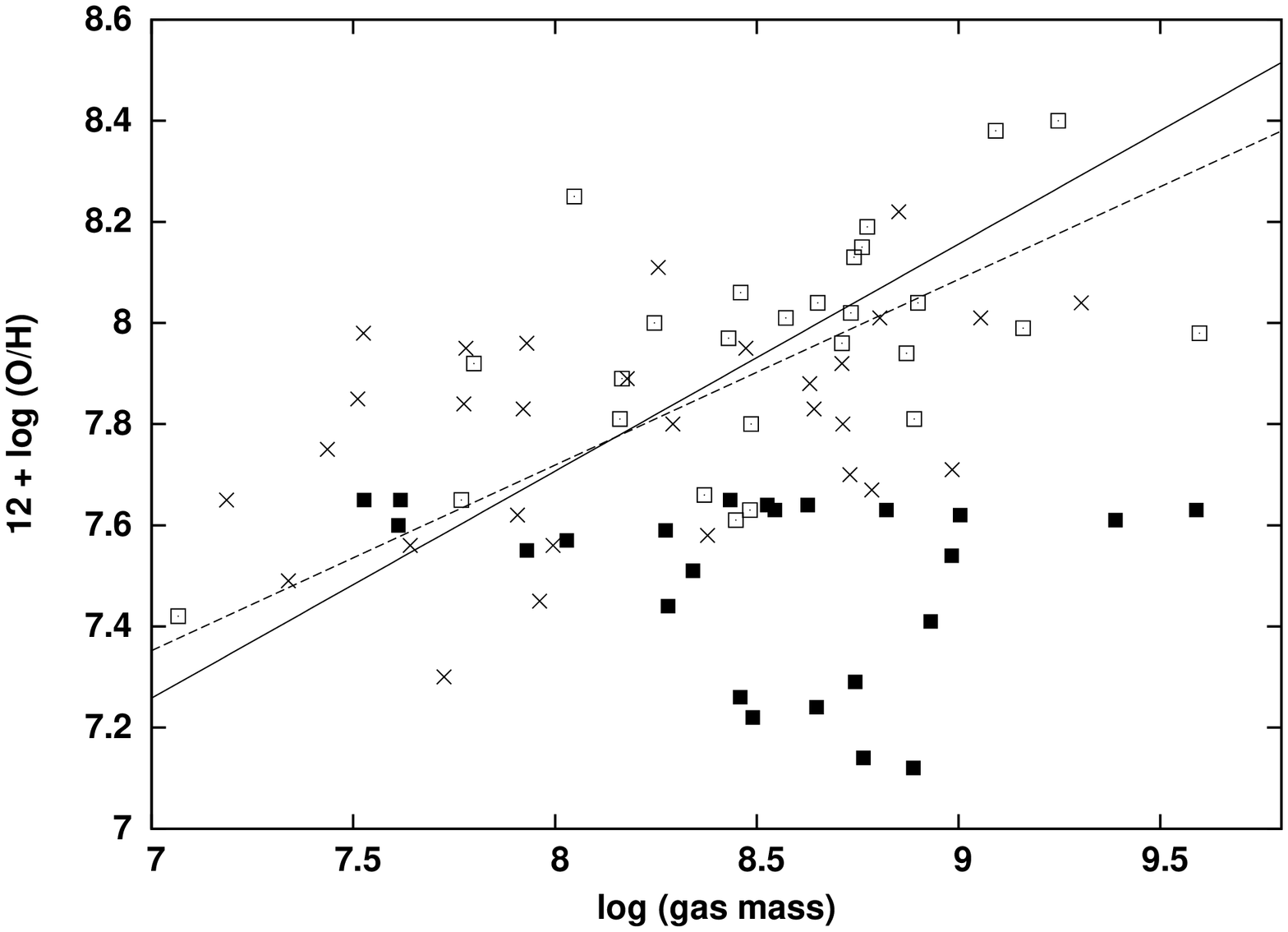}
\includegraphics[width=6.15cm,angle=270]{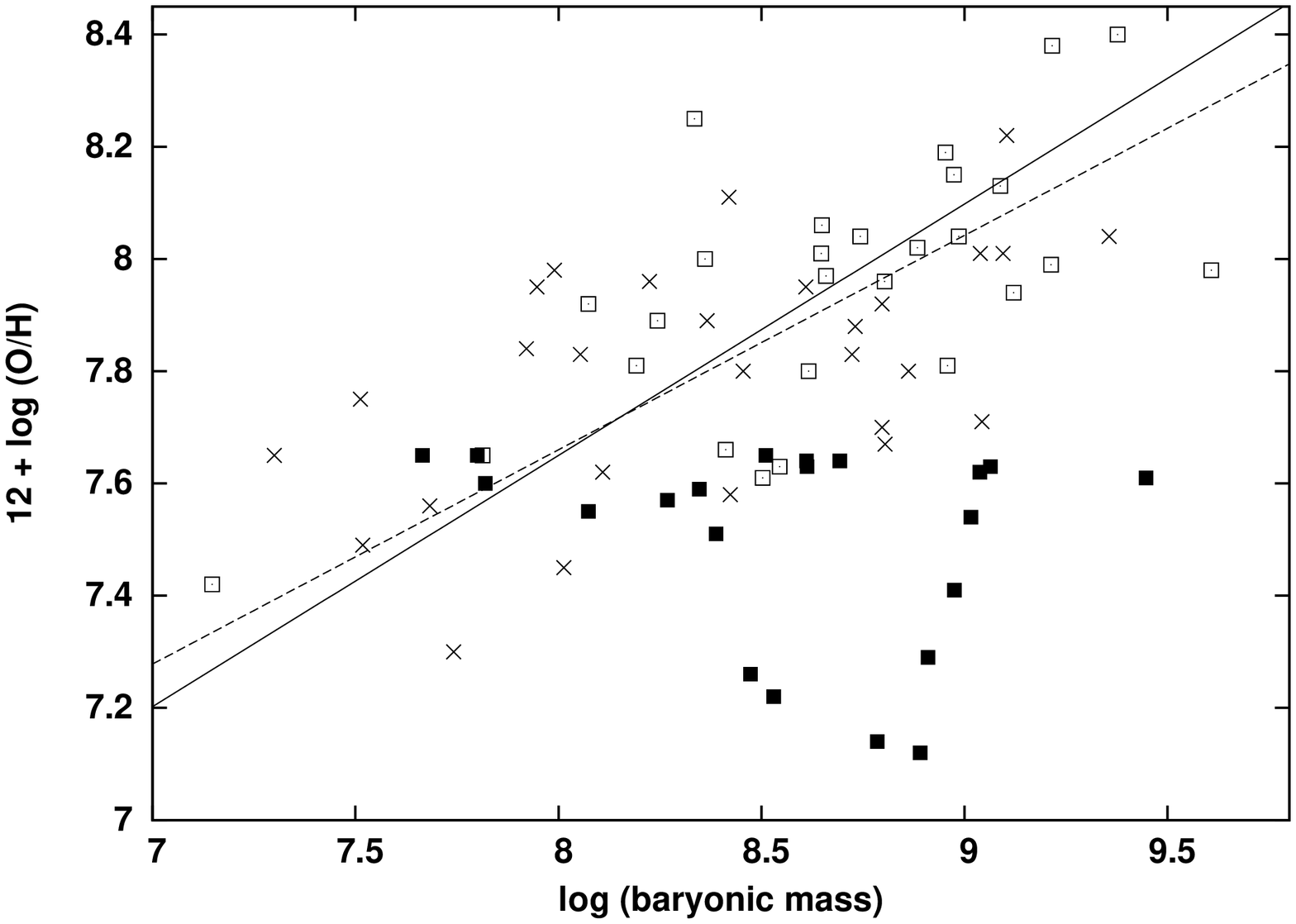}
\includegraphics[width=6.15cm,angle=270]{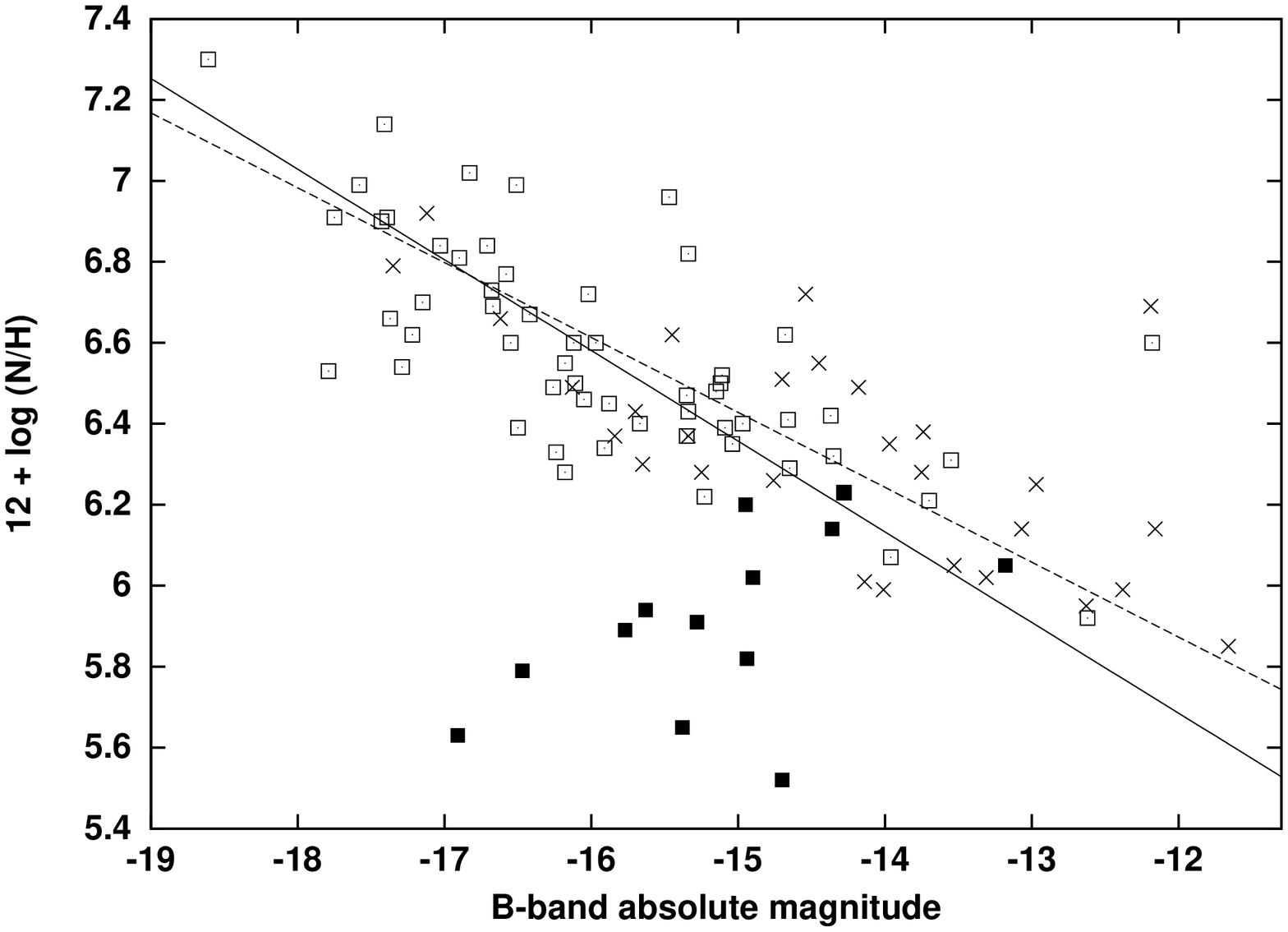}
\includegraphics[width=6.15cm,angle=270]{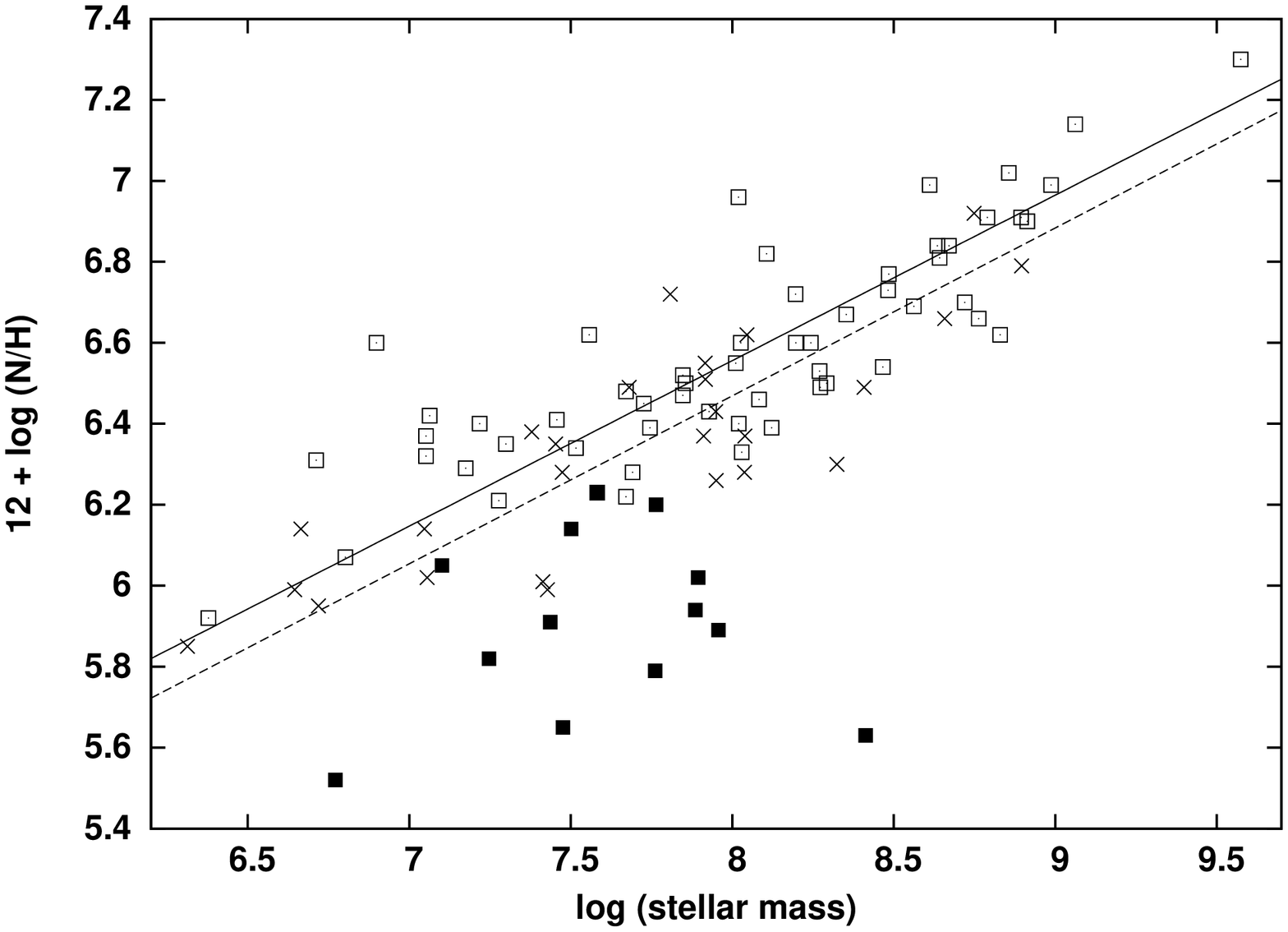}
\caption{The derived correlations of {\it B}-band luminousity 
{\bf (top left)}, stellar mass {\bf (top right)}, gas mass 
{\bf (middle left)}, baryonic mass {\bf (middle right)} with 
oxygen abundance, and {\it B}-band luminousity {\bf (bottom left)} 
and stellar mass {\bf (bottom right)} with nitrogen abundance, for 
BCGs (solid) and dIs (dashed) are shown. The BCG, dI 
and XMD BCG data are represented by empty squares, crosses and 
filled squares, respectively.}
\label{fig:lz}
\end{figure*}

The correlations of oxygen abundance [12~+~log(O/H)] with {\it B}-band 
absolute magnitude (L--Z), stellar mass (M$_{*}$--Z), gas mass (M$_{gas}$--Z) 
and baryonic mass (M$_{bary}$--Z) for BCG and dI samples are given in 
Table~\ref{tab:lztab}. We also provide the correlations of nitrogen abundance 
with {\it B}-band absolute magnitude (L--N) and stellar mass (M$_{*}$--N), for 
both samples, in the same Table. These correlations, along with the data points 
themselves, are shown in Fig.~\ref{fig:lz}. To the best of our knowledge, 
this is the first derivation of the mass--metallicity relations for BCGs.
We find that the L--Z and M--Z correlations for BCGs are dIs are, by and large, 
similar, except that the L--Z relation for BCGs is slightly steeper than that 
of dIs. Also, the two L--Z relations are slightly offset, such that for a given 
metallicity, BCGs are slightly more luminous than dIs. The same trends, in L--Z 
relation, were 
noted earlier by Lee et al. (2004) for their sample of \HII\ galaxies, who 
suggested that the ongoing star burst has temporarily enhanced the luminousity 
of these galaxies compared to dIs. Consistent with this suggestion,
the M$_{*}$--Z correlations for dIs and BCGs are much better matched. 
M$_{gas}$--Z and M$_{bary}$--Z relations for dIs and BCGs also match within 
their error bars. Nevertheless, there seems to be a suggestion (especially, 
from the M$_{*}$--Z plots and data points) that dIs are slightly 
metal-poor than BCGs, for a given mass. \cite{vaduvescu} also made a 
similar observation for their dI sample and a small sample of Virgo Cluster 
blue compact dwarf (BCD) galaxies, and found that for a given metallicity, 
BCD galaxies have lower baryonic masses than dI galaxies. The correlations 
using nitrogen, as a metallicity tracer, are similar to those derived from the 
oxygen abundance. The published L--Z and M--Z relations of dIs (viz., Lee et 
al. 2003; van Zee \& Haynes 2006; Lee et al. 2006; Vaduvescu et al. 2007) and 
L--Z relation of \HII\ galaxies (Lee et al. 2004) are quite 
close to those we derive from our samples.

\begin{figure}
\includegraphics[width=6.15cm,angle=270]{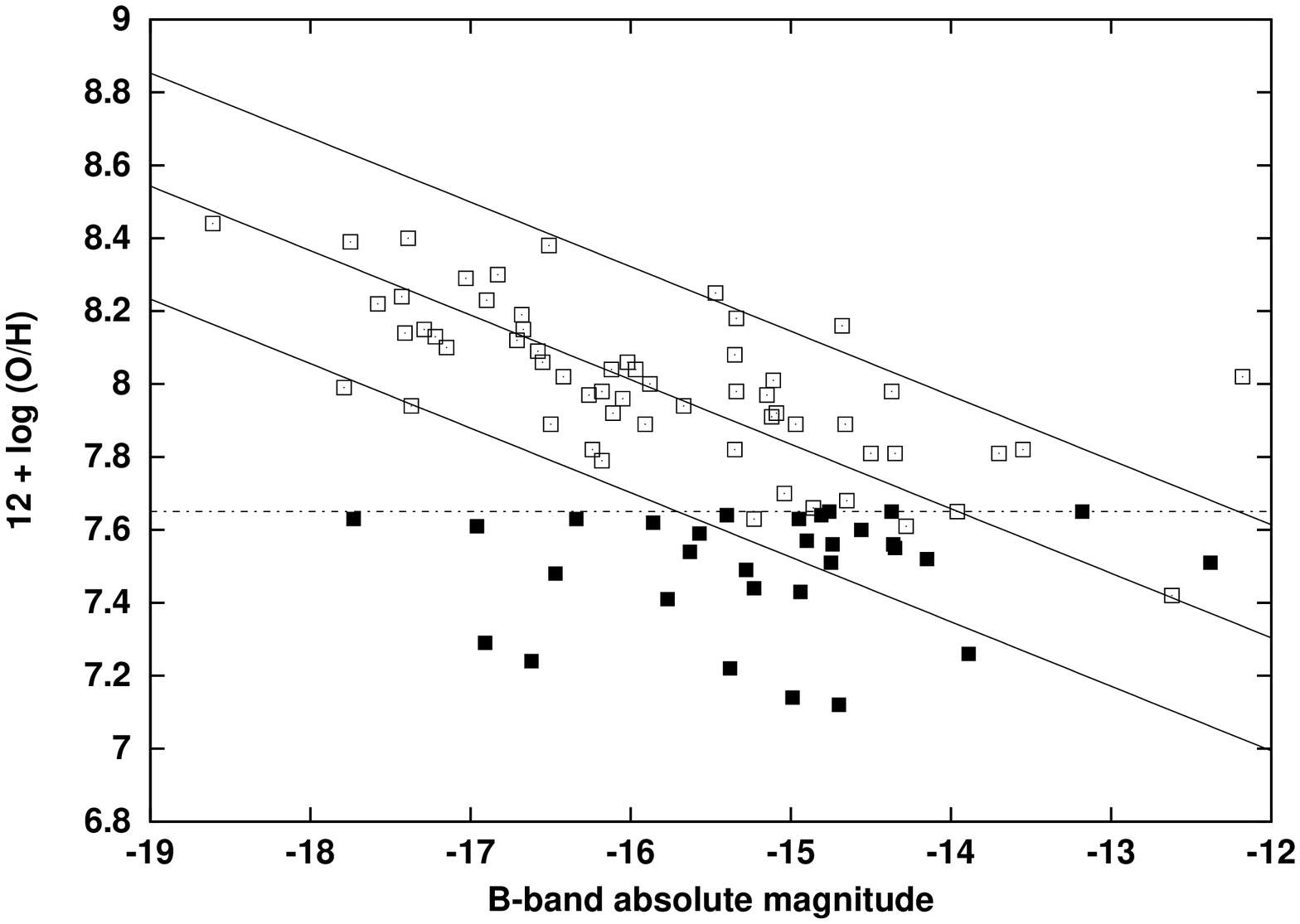}
\caption{BCG (empty squares) and XMD (filled squares) galaxies data points 
are plotted in the L--Z plane. The best-fitting regression line to the BCG galaxies 
is shown, along with the 95 per cent confidence interval about this line. 
The dividing line between XMD and non-XMD galaxies [12~+~log(O/H)~=~7.65] 
is also shown.}
\label{fig:xmd_new}
\end{figure}

As for our XMD sample, while some of the low-luminousity galaxies lie along the 
derived L--Z and M--Z relations, other XMD galaxies are deviant, and more so at 
larger masses and/or lower metallicities (Fig.~\ref{fig:lz}). In order to find 
the statistical significance of these deviations, we have drawn 95~per~cent 
confidence interval about the L--Z relation for BCGs, along with data 
points of XMD galaxies, as shown in Fig.~\ref{fig:xmd_new}. Note that the 
confidence interval, around the regression line, was computed using standard 
formulae (see, e.g., Montgomery, Peck \& Vining 2006). While the 
regression line, itself, was computed using the Geometric Mean Regression 
(GMR), this confidence interval was calculated assuming that 
12~+~log(O/H) is the independent variable. However, the width of 
the confidence interval, computed as above, should also be a sufficient 
measure of the confidence interval around the GMR. 

 Fig.~\ref{fig:xmd_new} also shows the metallicity 
criterion (viz., 12~+~$\log$(O/H)~$\le$~7.65) used for defining XMD galaxies. 
We note that for all BCGs with M$_{B}$~$>$~--15.7, a metallicity 
of 12~+~$\log$(O/H)~=~7.65 lies within the 95~per cent confidence interval 
about the L--Z relation. Conversely, BCGs brighter than M$_B$~$\sim$--15.7 can 
have 12~+~$\log$(O/H)~=~7.65, but still be outliers with respect to the L--Z 
relation. The lower bound to the 95~per cent confidence interval is given by 
Z~=~--0.177~M$_B$~+~4.87, and only those galaxies which have metallicities 
below this are `extremely metal-deficient' in a statistically significant 
manner. We note that the choice of the 95~per cent confidence interval is 
somewhat arbitrary, but suggest it, none the less, as a reasonable working 
definition. Of our sample of XMD galaxies, we find that more than half 
(17 of the 31) are classified as XMD by this definition, and in fact, nine 
of them lie outside the 99.5~per cent confidence interval about the L--Z 
relation. Further, it remains true that galaxies with 12~+~log(O/H)~$\le$~7.65 
are significantly more metal-deficient than the gas in the solar neighbourhood, 
and in this sense, they can still be regarded as `extremely metal-deficient'.

\begin{table*}
\label{tab:lztab}
\begin{tabular}{ll}
\hline
12~+~log(O/H)~=~(--0.177~$\pm$~0.016)~M$_{B}$~+~(5.18~$\pm$~0.25) & 58 BCGs \\
12~+~log(O/H)~=~(--0.142~$\pm$~0.019)~M$_{B}$~+~(5.76~$\pm$~0.28) & 32 dIs \\
12~+~log(O/H)~=~(0.313~$\pm$~0.025)~log(M$_{*}$)~+~(5.50~$\pm$~0.20) & 59 BCGs \\
12~+~log(O/H)~=~(0.336~$\pm$~0.038)~log(M$_{*}$)~+~(5.22~$\pm$~0.29) & 30 dIs \\
12~+~log(O/H)~=~(0.449~$\pm$~0.070)~log(M$_{gas}$)~+~(4.12~$\pm$~0.60) & 27 BCGs \\
12~+~log(O/H)~=~(0.367~$\pm$~0.060)~log(M$_{gas}$)~+~(4.78~$\pm$~0.50) & 30 dIs \\
12~+~log(O/H)~=~(0.448~$\pm$~0.065)~log(M$_{bary}$)~+~(4.07~$\pm$~0.56) & 27 BCGs \\
12~+~log(O/H)~=~(0.382~$\pm$~0.060)~log(M$_{bary}$)~+~(4.60~$\pm$~0.26) & 28 dIs \\ 
12~+~log(N/H)~=~(--0.224~$\pm$~0.019)~M$_{B}$~+~(3.00~$\pm$~0.32) & 56 BCGs \\
12~+~log(N/H)~=~(--0.185~$\pm$~0.021)~M$_{B}$~+~(3.65~$\pm$~0.32) & 28 dIs \\
12~+~log(N/H)~=~(0.409~$\pm$~0.028)~log(M$_{*}$)~+~(3.28~$\pm$~0.22) & 56 BCGs \\
12~+~log(N/H)~=~(0.415~$\pm$~0.044)~log(M$_{*}$)~+~(3.15~$\pm$~0.34) & 26 dIs \\
\hline
\end{tabular}
\caption{Derived L--Z and M--Z correlations}
\label{tab:lztab}
\end {table*}

\begin{figure}
\includegraphics[width=6.1cm,angle=270]{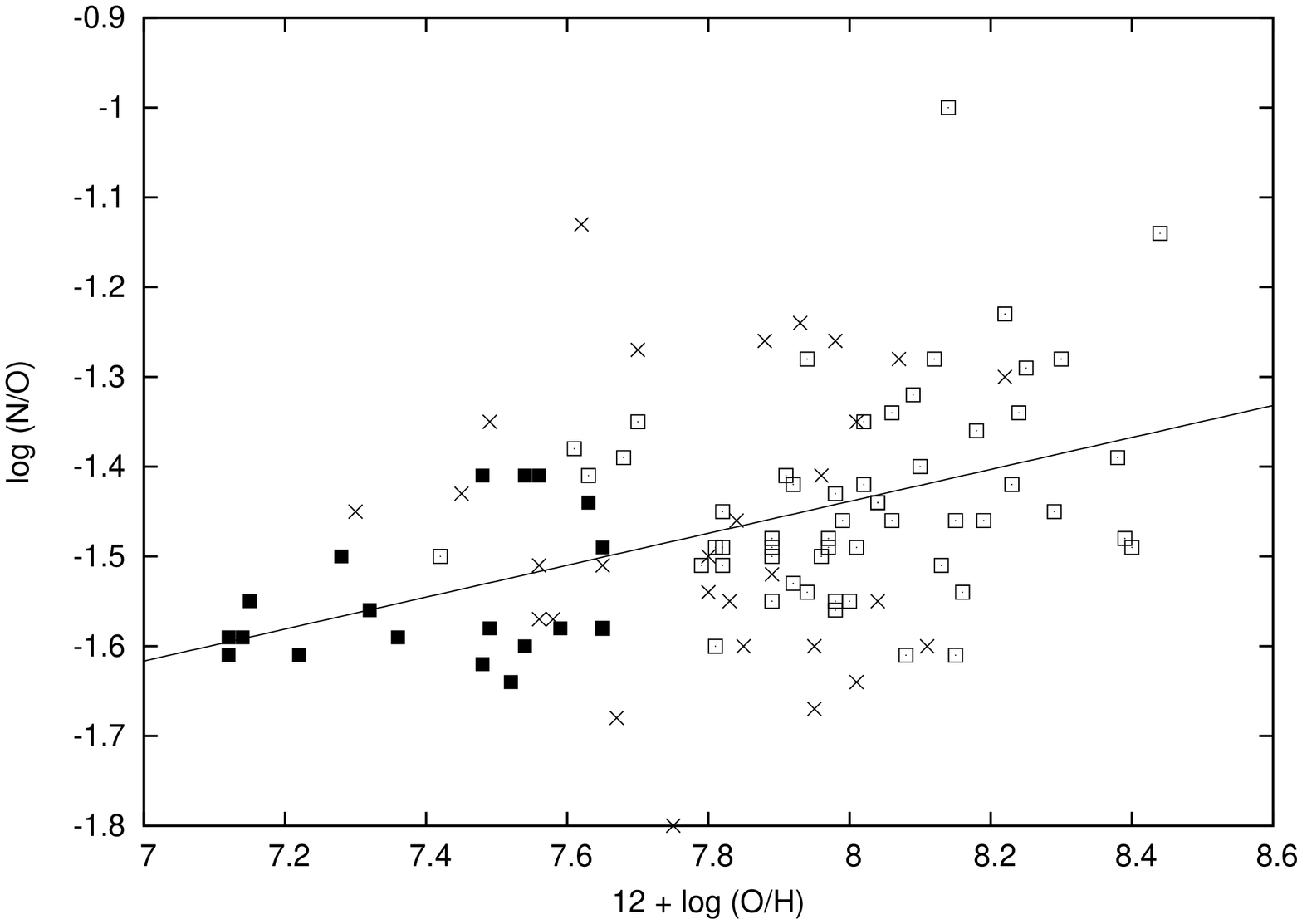}
\caption{The nitrogen to oxygen ratio [log(N/O)] as a function of the
oxygen abundance 12~+~log(O/H). The symbols are BCGs (empty squares), 
dIs (crosses) and XMD BCGs (filled squares). For XMD galaxies in which 
individual \HII\ region metallicities are known, each is shown as
a separate point. These are taken from Izotov et al. (2005), Izotov 
et al. (2009), Izotov et al. (2006a), Izotov et al. (2004), Izotov \& Thuan 
(1999), Izotov, Thuan \& Lipovetsky (1997), Guseva et al. (2003a), Guseva 
et al. (2003b), Lee et al. (2004) and Melbourne et al. (2004).   
The best-fitting line (i.e., linear regression) to a combined sample, 
including BCGs in our 
normal sample and \HII\ regions of XMD BCGs (i.e., both empty and 
filled squares) is plotted.}
\label{fig:N/OvsO}
\end{figure}

The nitrogen-to-oxygen abundance ratio [log(N/O)] is plotted against oxygen 
abundance in Fig.~\ref{fig:N/OvsO}. We find a moderate correlation between 
log(N/O) and 12~+~log(O/H) for a combined sample, consisting of our BCG 
and XMD BCG samples (correlation coefficient is 0.51~$\pm$~0.07).
While it is well-known that these quantities correlate at
high metallicities [12~+~log(O/H) above 8.3, see, e.g., Henry, Edmunds \& 
Koppen (2000)] it is interesting that we find a correlation, even though 
most of our galaxies are below this metallicity. However, consistent 
with the findings in Izotov et al. (2006a), the N/O ratio does show signs 
of flattening at low metallicities, i.e., 12~+~log(O/H) below 7.4.

We have, so far, been considering the metallicity of a galaxy, which, in
some sense, is a measure of the amount of its past star formation activity. 
Specifically, in a closed-box model (assuming, further, that the delay 
between star formation and the release of metal-enriched gas is negligible,
and that the gas is well-mixed) the metallicity of a galaxy is related to
its gas mass fraction ($\mu$$_{gas}$) and the `chemical yield' (p) by 

\begin{equation}
{\rm Z} = {\rm p} \ln(1/\mu_{\rm gas}) 
\label{eq:p}
\end{equation}
[where the yield (p) is defined as the amount of metals produced, per unit mass 
that is locked up in stars, see, e.g., Binney \& Merryfield (1998)]. Galaxies 
may not evolve as closed boxes, either because they gain (pristine) material 
due to infall, or lose (enriched) gas through galaxy winds. The quantity p, in 
Equation~\ref{eq:p}, then represents the effective chemical yield (\peff), 
\begin{equation}
\peff = -Z/\ln(\mu_{\rm gas}), 
\label{eq:peff}
\end{equation}
of a galaxy. 

\begin{figure}
\includegraphics[width=6.0cm,angle=270]{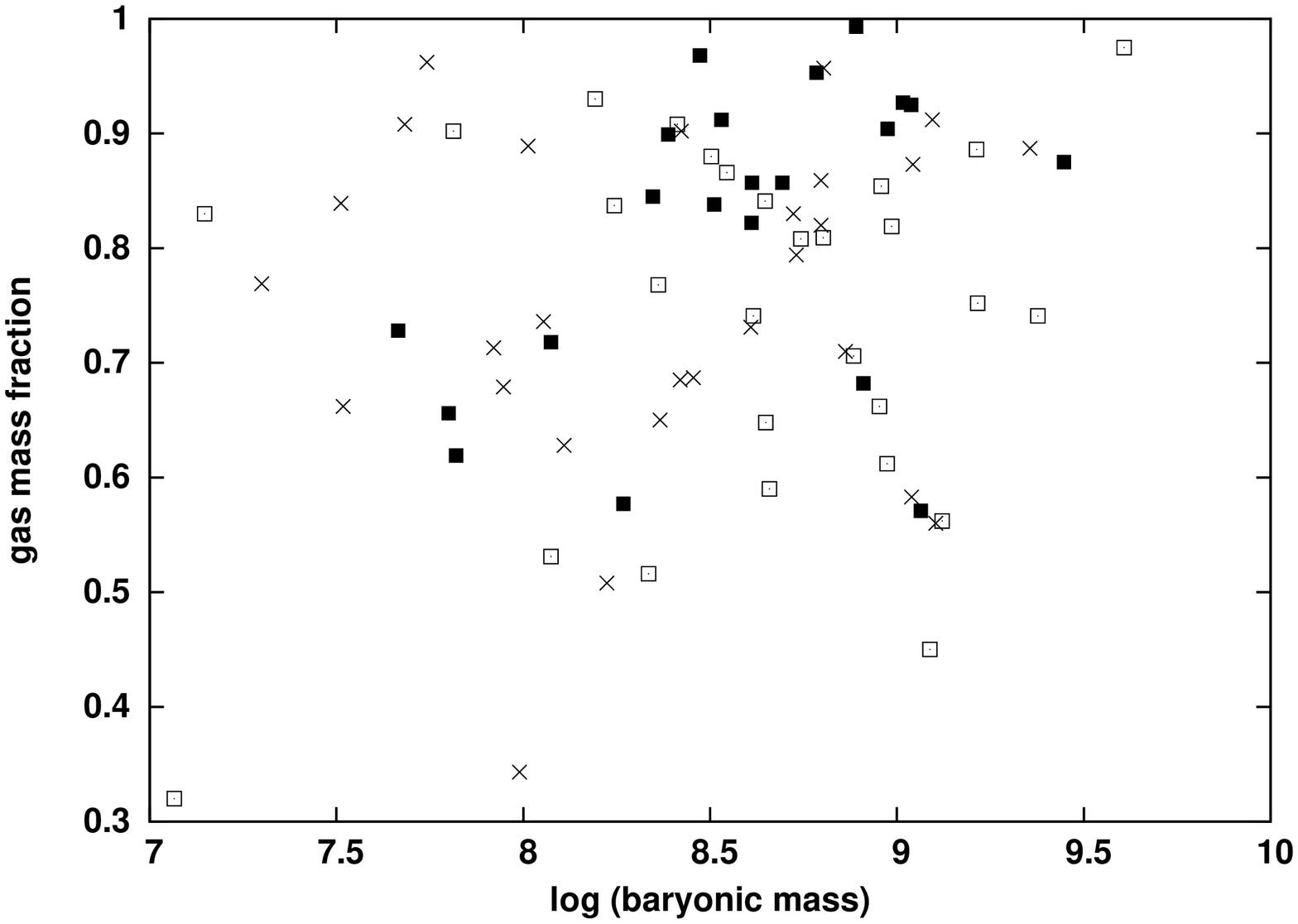}
\includegraphics[width=6.0cm,angle=270]{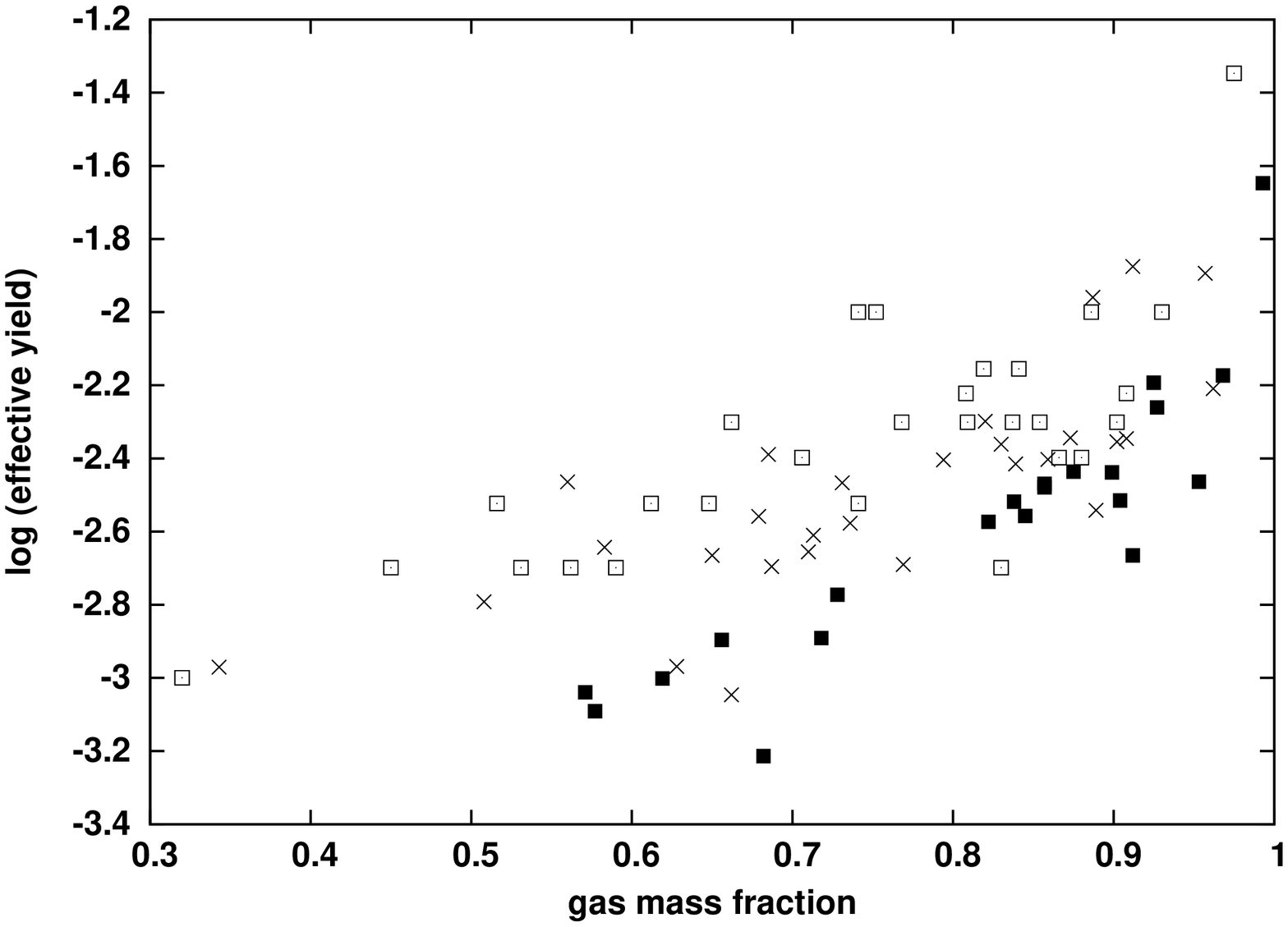}
\includegraphics[width=6.0cm,angle=270]{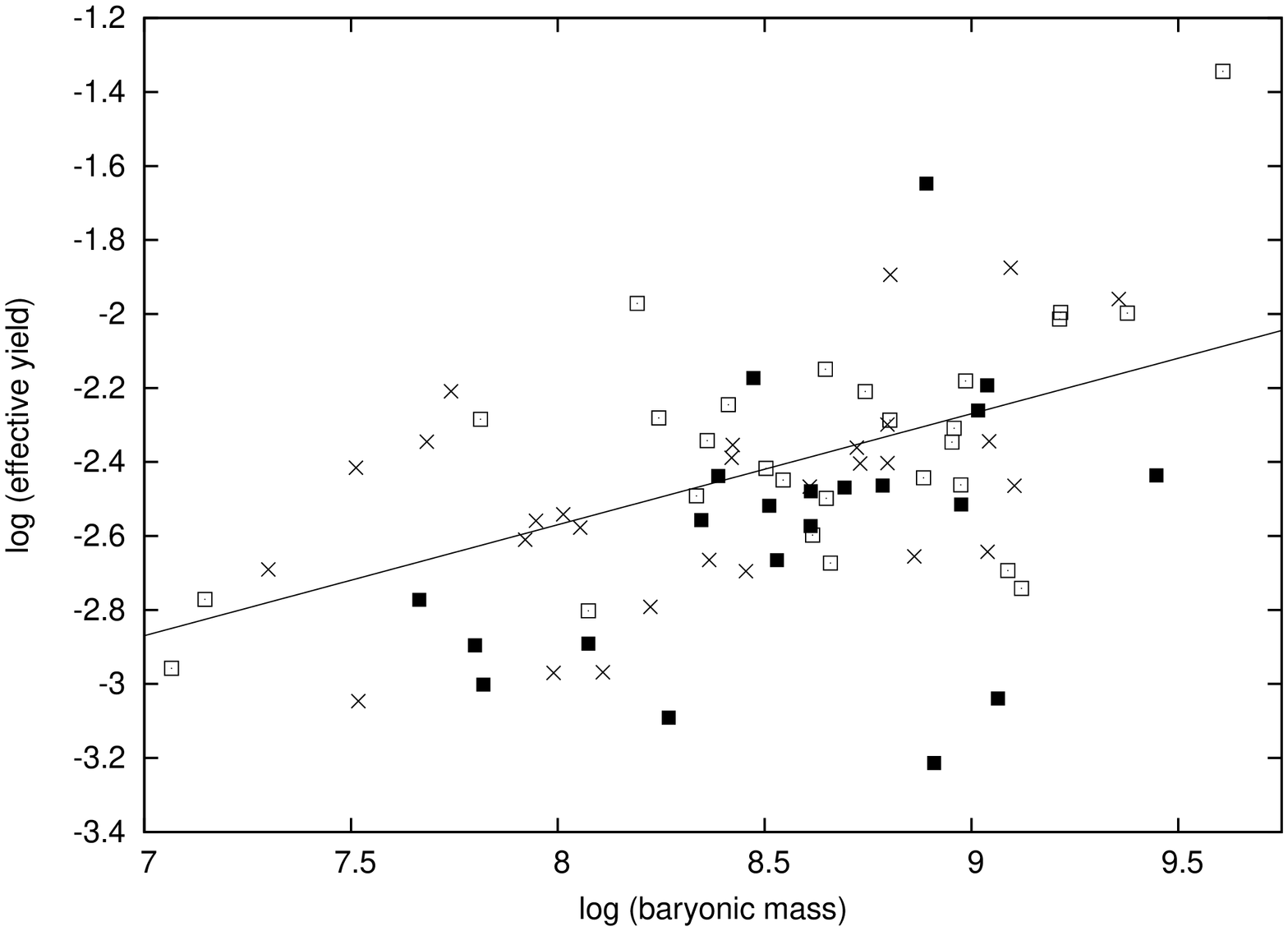}
\caption{Gas mass fraction is plotted against baryonic mass {\bf (top)}, and 
effective yield is plotted against gas mass fraction {\bf (middle)} and 
baryonic mass {\bf (bottom)} for BCGs (empty squares), XMD BCGs (filled 
squares) and dIs (crosses). The best-fitting line (i.e., linear regression) to 
the combined BCG and dI 
sample (XMD BCGs were not included) is shown in the bottom panel.}
\label{fig:pmumbary}
\end{figure}

As can be seen from the definition, \peff\ depends on the metallicity and
the gas mass fraction. In Fig.~\ref{fig:pmumbary}(top), we show the gas
mass fraction of our sample of galaxies. As can be seen, XMD galaxies
tend to be somewhat gas-richer than BCG or dI galaxies. The sample
mean \fgas\ of the three samples are 0.74$\pm$0.03 (BCG), 
0.75~$\pm$~0.03 (dI) and 0.82~$\pm$~0.03 (XMD). Interestingly, the most 
discrepant XMD galaxies (i.e., those which lie below the 95~per cent 
confidence interval determined above) tend to have the highest gas fractions. 
The fact that XMD 
galaxies have larger gas mass fractions means that they have converted
a smaller fraction of their baryonic mass into stars, and as such, one
would expect them to have lower metallicities. If this was the only
reason for their low metallicities, then XMD galaxies would have  
effective yields comparable to those of other gas-rich dwarf galaxies. 
Fig.~\ref{fig:pmumbary}(middle) shows the effective yield as a function 
of the gas mass fraction. One can see a clear separation between XMD and the 
other dwarf galaxies -- at the same gas mass fraction, XMD
galaxies have systematically lower effective yields. So clearly, there
is some other phenomenon at work -- perhaps, XMD galaxies have lost 
more of their metals via preferential outflow of metal-enriched
gas? One would expect that the outflow of metal-enriched gas would 
increase with decreasing depth of the dark-matter halo potential of a galaxy.
Following Tremonti et al. (2004), we use the baryonic mass as a proxy for 
the mass of dark matter [since the two are expected to be tightly correlated, 
because of the baryonic Tully-Fisher relation (McGaugh 2005)]. 
Fig.~\ref{fig:pmumbary}(bottom) shows the effective yield as a function 
of the baryonic mass. As can be seen, one does indeed see an increase 
in the effective yield with baryonic mass (the correlation coefficient 
computed for the BCG and dI samples is 0.54~$\pm$~0.10, with the error bar 
computed via bootstrap resampling). However, what is interesting to note is 
that {\it at the same baryonic mass}, XMD galaxies have systematically lower 
effective yields than BCG or dI galaxies. Clearly again, some other
process is at work in XMD galaxies.

   In this context, it is interesting to note that \HI\ studies of 
XMD BCGs (van Zee et al. 1998; Chengalur et al. 2006; Ekta et al. 2006, 2008, 
2009; Ekta \& Chengalur 2010) have shown that interactions/mergers 
are common in them. Recall that a fundamental assumption in deriving
the effective yield is that the gas is {\it well-mixed}, i.e., any metals 
formed during star formation are uniformly mixed throughout the gaseous 
envelope. This is unlikely to be true in extremely gas-rich dwarf galaxies, 
where the gas disc is substantially more extended than the stellar disc 
(Begum et al. 2008). This means that for these galaxies, if the effective yield 
is derived assuming that there is perfect mixing of all the gas, it will lead 
to an overestimate of the true yield. Note that in Fig.~\ref{fig:pmumbary}
(middle and bottom), the observed effective yields for BCGs and dI galaxies 
exceed the theoretically-computed closed-box yield [viz., an 
effective yield of 0.0074, i.e., log(p$_{eff}$) of --2.13 (Meynet \& Maeder 
2002)] while those of XMD galaxies are generally lower than this. In the
absence of the infall of chemically-enriched gas, the closed-box
yield is the maximum possible effective yield. Given the uncertainties
in the theoretical calculations of the chemical yield and measurement
errors, this can not be regarded as definitive evidence that the
effective yields of gas-rich galaxies are overestimated if one
includes the entire gas mass in the \peff\ calculation, but it
is certainly consistent with this idea.
 
\section{DISCUSSION \& SUMMARY}                        
\label{sec:dis}

There are, in principle, three possible reasons for the low metallicity of an 
XMD galaxy, viz., (i)~low star formation efficiency, (ii)~an outflow of 
metal-enriched gas, or (iii) an inflow of metal-poor gas. 
In disentangling these possibilities, it is useful to look at the effective 
chemical yields of the galaxies. Brooks et al. (2007) argue that
low star formation efficiency of low-mass galaxies is primarily responsible
for the M--Z relation. On the other hand, if the low metallicity were merely 
a consequence of low star formation efficiency, then all galaxies would 
have the same effective yield, which, as we saw in the previous Section, 
is not the case. Further, as can be readily shown (see, e.g., Dalcanton 2007)
inflow of metal-poor gas can not substantially lower the effective yield 
of extremely gas-rich galaxies -- for such galaxies, the effective yield 
asymptotes to a constant value (M$_{\rm Z}$/M$_{\rm star}$) independent 
of the gas mass. Since most of our XMD galaxies are very gas-rich, it seems 
unlikely that their low metallicities are due to inflow of pristine 
material. Another possibility is an outflow of metal-enriched gas. 

Numerical simulations  (e.g., Maclow \& Ferrara 1999) suggest that dwarf 
galaxies can lose a substantial fraction of their metals due to outflows driven 
by supernova explosion. The amount of metal loss, however, depends on both, 
the supernova rate and the spatial and temporal correlation of the explosions, 
and other simulations (e.g., Fragile, Murray \& Lin 2004) indicate a much 
lower metal loss from dwarf galaxies. On the observational front, Tremonti et 
al. (2004) argue that the metal loss does, indeed, anti-correlate with the 
depth of the gravitational potential well (or equivalently, the effective yield 
correlates with the total baryonic mass) as would be expected in the outflow 
scenario. As we saw in the previous Section, a similar trend is seen in our 
samples of dwarf galaxies. On the other hand, Fig.~\ref{fig:pmumbary} also 
shows that {\it at the same baryonic mass}, XMD galaxies have lower effective 
yields than other gas-rich dwarf galaxies. It, thus, appears that some process, 
particular to XMD galaxies, is responsible for their low metallicity. 

   Motivated by the observation that many of the XMD galaxies, with detailed
\HI\ observations, are undergoing interactions (e.g., Ekta et al. 2008),
we propose that (other than being gas-richer, as seen in the previous 
Section) the mechanism responsible for the metal-deficiency of XMD 
galaxies is {\it a better mixing} of the ISM, as compared to the isolated, 
gas-rich dwarf galaxies. A similar argument was presented by Peeples, Pogge \& 
Stanek (2009), who found that metal-poor outliers from the M--Z relation of 
star-forming galaxies have disturbed morphologies. They argued that this is a 
consequence of tidally-induced inflow of gas to the galaxies' central regions, 
where low-metallicity gas, from large galactocentric radii, dilutes the 
central, metal-rich gas. A similar result, as that of the XMD galaxies being 
preferentially found in the interacting pairs, was also found by Kewley, 
Geller \& Barton (2006).
They show that at the same luminousity, close galaxy pairs have systematically 
low metallicity, compared to wide pairs and field galaxies. Further, 
Lee et al. (2004) found that galaxies with disturbed morphologies are offset 
towards higher luminousity and low metallicity in L--Z plane, and argue that
this could be a consequence of the mixing of outer, metal-deficient gas with
the ISM surrounding the central star-forming regions. 

Numerical simulations, presented by Bekki (2008), show the formation of BCD 
galaxies from merging between very gas-rich dwarf galaxies. 
They also show that since new stars can be formed in the centre of BCD 
galaxies, from the gas transferred from the outer part of the merger 
progenitors, new stars can be very metal-poor. More recently, Rupke, Kewley \& 
Barnes (2010) presented an analysis of simulations of mergers, which confirm 
that nuclear metallicity underabundances, observed in interacting disc 
galaxies, are due to merger-driven inflow of low-metallicity gas from the 
outskirts of merging galaxies. A subsequent analysis of gas-rich 
interaction and mergers by Montuori et al. (2010) shows a strong trend between 
the star-formation rate and dilution of metals in the nuclear region, due to 
inflows of metal-poor gas from the outer regions of merging discs, which fuels 
the intense star formation and lowers the metallicity.   
Finally, additional support to the idea, that the ISM in XMD 
galaxies is well-mixed, comes from Far Ultraviolet Spectroscopic Explorer 
(FUSE) observations of BCD galaxies which show that, in general, the neutral 
gas is more metal-poor as compared to the ionised gas in the \HII\ regions. In 
the most metal-deficient galaxies, however, the two are comparable 
(Lebouteiller et al. 2009 and references therein). 

 It is worth emphasizing that in the scenario being suggested here, it
is not so much that XMD galaxies have low chemical yields, as that the
effective yields of isolated, gas-rich dwarf galaxies are overestimated. 
This overestimate is because the gas in the outer parts of the galaxy
(which generally lies well outside the stellar disc, and does not 
participate in star formation) is, none the less, included in the calculation 
of effective yield. Better mixing of the gas, however, would not only lower
the metallicity in the central star-forming regions, but also bring the
galaxy closer to the well-mixed assumption in the closed box model, 
and hence, decrease the overestimation of the yield. We can now return 
to one of the questions that we had set ourselves at the start of this 
paper, viz., do XMD galaxies represent a separate population, or are they 
merely the low-metallicity tail of the metallicity distribution of galaxies?  
The conclusion that we reach is that they do indeed seem to be a separate 
population, one in which tidal interactions have led to a lower metallicity via 
a better mixing of the ISM.

To summarise, we try to address the question that whether the 
metallicities of XMD galaxies are consistent with those expected from 
the luminousity--metallicity and mass-metallicity relations of other
gas-rich dwarf galaxies, or are XMD galaxies discrepant.
To answer this, we consider the properties of three samples of galaxies, 
viz., a sample of XMD galaxies and comparison samples of BCGs and 
dI galaxies. We enlist our conclusions below.

\begin{enumerate}
\item For sufficiently low-luminousity (i.e., $M_B$~$>$~--15.7) galaxies, 
      a metallicity of 12~+~log~(O/H)~$\sim$~7.65 (i.e., the threshold
      metallicity for XMD galaxies) is consistent with what one would 
      expect from the luminousity--metallicity relation. 

\item The lower bound to the 95~per cent confidence interval, around the
      L--Z relation, is given by Z~=~--0.177~M$_B$~+~4.87. We suggest that this
      would be a more appropriate definition for XMD galaxies. As per 
      this definition, 17 of the 31 galaxies, in our sample, would be 
      identified as XMD.

\item As shown in previous work, we find that the L--Z relation of BCGs 
      is shifted to higher luminousities at a given metallicity, compared 
      to dIs. This has been considered as an evidence for the fact that the 
luminousity of BCGs is temporarily enhanced due to ongoing starburst. 
Consistent with this
      suggestion, we find that the M--Z relations of our BCG and dI 
       samples are better matched.  

\item The effective chemical yields of XMD BCGs are systematically lower 
      than BCGs and dI galaxies, with similar gas fractions and baryonic 
masses. 
      We suggest that this is because of better mixing of the ISM in the 
      case of XMD galaxies, as compared to other gas-rich dwarf galaxies. 
Motivated by the observation that many XMD galaxies are interacting (Ekta et 
al. 2008) we suggest that the better mixing is due to tidal interactions.
We propose that XMD galaxies deviate from the L--Z and M--Z relations because 
of a combination of having low effective chemical yields and higher 
gas fractions. 
\end{enumerate} 

~\\

{\bf ACKNOWLEDGMENTS} \\
We thank the anonymous referee for useful comments, which helped 
in increasing the clarity of the paper. 
This research has made use of the NASA/IPAC Extragalactic Database (NED) 
which is operated by the Jet Propulsion Laboratory, California Institute 
of Technology, under contract with the National Aeronautics and Space 
Administration. Funding for the SDSS and SDSS-II has been provided by the 
Alfred P. Sloan Foundation, the Participating Institutions, the National 
Science Foundation, the U.S. Department of Energy, the National Aeronautics 
and Space Administration, the Japanese Monbukagakusho, the Max Planck Society, 
and the Higher Education Funding Council for England. The SDSS Web Site is 
http://www.sdss.org/. 
The SDSS is managed by the Astrophysical Research Consortium 
for the Participating Institutions. The Participating Institutions are the American 
Museum of Natural History, Astrophysical Institute Potsdam, University of Basel, 
University of Cambridge, Case Western Reserve University, University of Chicago, 
Drexel University, Fermilab, the Institute for Advanced Study, the Japan Participation 
Group, Johns Hopkins University, the Joint Institute for Nuclear Astrophysics, the Kavli 
Institute for Particle Astrophysics and Cosmology, the Korean Scientist Group, the Chinese 
Academy of Sciences (LAMOST), Los Alamos National Laboratory, the Max-Planck-Institute for 
Astronomy (MPIA), the Max-Planck-Institute for Astrophysics (MPA), New Mexico State 
University, Ohio State University, University of Pittsburgh, University of Portsmouth, 
Princeton University, the United States Naval Observatory, and the University of Washington.

\label{lastpage}

\end{document}